# Modeling the Safety Effect of Access and Signal Density on Suburban Arterials: Using Macro Level Analysis Method

Yuan Jinghui [1,2] and Wang Xuesong [1,2]

(1. The Key Laboratory of Road and Traffic Engineering, Ministry of Education, Tongji University, 4800 Cao'an Road, Jiading District, Shanghai, 201804, China;

2. Jiangsu Province Collaborative Innovation Center of Modern Urban Traffic Technologies，SiPaiLou #2，Nanjing，210096，P.R. China)

**Abstract:** With rapidly increasing of the land development density along suburban arterials, much more irregular signal spacing appeared on suburban arterials, and high access density is commonly observed on suburban arterials. These issues tend to increase the risk of crash occurrence of arterials. By developing safety performance functions on road segments and intersections separately, the previous research analyzed the partial safety effects of the influence factors. In this study, Bayesian Conditional Autoregressive (CAR) models were developed at traffic analysis zone (TAZ) level for suburban arterials laid in suburban area in Shanghai. The model result showed that higher access and signal density tend to increase crash frequencies occurred on arterials. At this point, designing frontage roads paralleled to arterials to collect the access traffic instead of those intensive access could reduce crashes occurred on arterials.

## 1 INTRODUCTION

Rapidly development brings higher traffic demand along these suburban arterials accompanied by more accesses and intersections, which resulted in an increasing lateral disturbance to arterials. As for the signal spacing, much more non-uniform signal spacing appeared on suburban arterials. Besides, high access density is commonly observed on suburban arterials which leads to an increase of the lateral disturbance. These arterials are gradually beginning to act as an urban road. While suburban arterials were originally planned for longer distance trips, and the operation speed is higher than those on urban streets. There is no current authoritative guideline for regulating signal spacing and access density for arterials in the suburban area. All these issues tend to increase the risk of crash occurrence on arterials.

Previous studies on arterial safety have investigated safety by modeling road segment or intersection separately. The limitation of this approache is that they only estimates the partial effects of the influence factors on arterial safety. While the signal density and access density have safety impacts on both segments and intersections. Thus, an overall and in-depth analysis on the relationship between those factors and crash occurrence of arterials is needed. The purpose of this paper is to report the results of research conducted to perform a macro level safety modeling method applied at the Traffic Analysis Zone (TAZ) level to analyze the safety of arterial roadways. Try to analyze the full impacts of access density and signal density on arterial safety.

## 2 LITERATURE REVIEW

Several studies have been conducted on the safety analysis for arterials over the past years. The majority of these studies were conducted at the road segment level. There exist several different approaches for road segmentation, primarily including fixed length [1] and homogeneous segments [2]. As for intersection, the crash data were primarily collected from the intersection functional area [3][4]. However, since there still didn't exist a uniform method for the road segmentation or the intersection analysis scope identification. Different segmentations or identification would result in differences in the data allocation, which may lead to bias model estimation. Above all, those research could not estimate the full safety impacts of influence factors on arterials (including road segments and intersections).

Previous researchers have shown the significance of signal density and access density on crash occurrence. With respect to signal density, several studies have shown that more crashes occurred on those roadways with higher signal density [5]. With respect to access density, several studies have found that higher access density leads to more crashes [5].

During the previous research on macro level safety analysis, the most popular analysis unit was TAZ [6]. In this study, we aimed at the crash occurrence on suburban arterials at the TAZ level to evaluating the full impacts of specific variables on arterial safety. Conventional Generalized Linear Models (GLM) were developed under the assumption that all samples are independent, however, there are often spatial correlations across the crash data of neighboring zones. To address this limitation, several studies have been conducted by applying the Conditional Autoregressive (CAR) model [7].

Bayesian methods are preferred by many researchers to estimate model coefficients since Bayesian methods can deal with complex data structures effectively. At this point, spatial statistical models using Bayesian method have been applied extensively in traffic safety research [7].

## 3 DATA PREPARATION

This study was conducted on the suburban area of Jiading and Baoshan District in Shanghai. The modeling data were prepared from three aspects: TAZ delineation, extraction of road features and crash allocation of suburban arterials for each delineated TAZ.

### 3.1 TAZ delineation and data extraction

According to the rules of TAZ delineation, overlay the five shapes of Jiading and Baoshan District (outside the Outer Ring) using ArcGIS®, including river shape, regionalism shape, highway shape, town area shape and railway route shape to form an initial TAZ boundary shape. Eventually, the 202 delineated TAZs are presented by the hidden line in Figure 1.

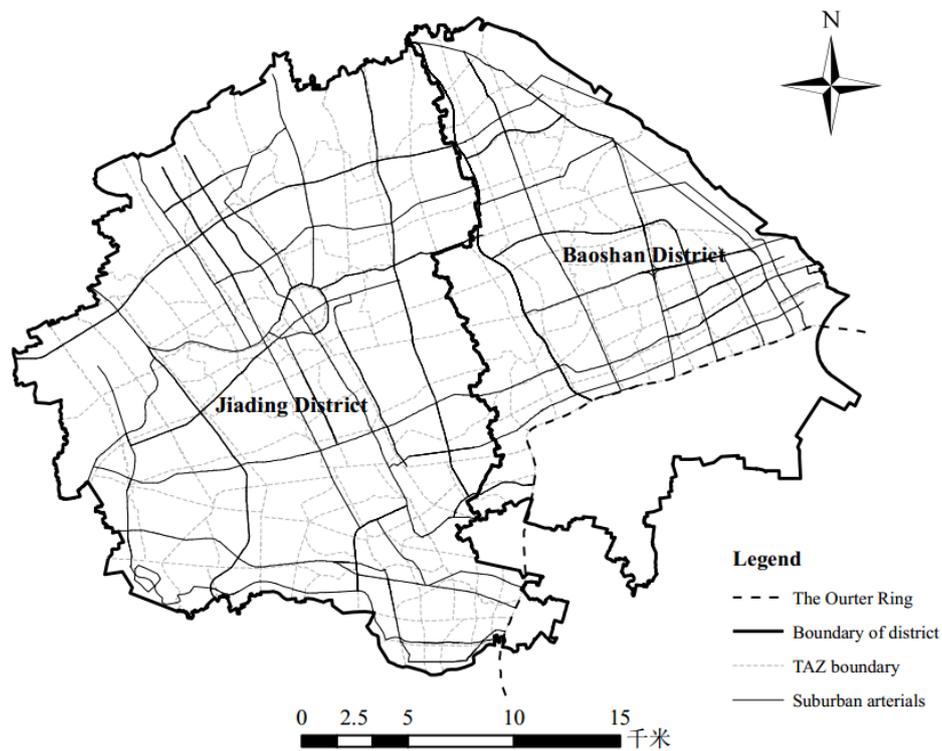

**Figure 1. Suburban arterials in Jiading and Baoshan District (outside the Outer Ring).**

The land use properties distribution of Jiading and Baoshan District were acquired from the website of the Municipal Land Administration Bureau of Jiading and Baoshan District. All the land use types were classified into 7 classes: industrial land, commercial land, educational land, technical land, residential land, greenspace land and agricultural land. Descriptive statistics of the land use percentage is shown in

Table **2**.

Road features mainly include the arterial length, signal density and access density along arterials. As to the arterial length, it was calculated in the total mileage of arterials in every TAZ using the function of spatial join in ARCGIS®. For the signal density and access density, calculate the number of signal intersections and access along the arterials within each TAZ, and divided them by the arterial length of each TAZ, then the corresponding signal density and access density could be generated. A descriptive statistics of these variables are listed in

Table **2**.

Table 1. Proportion Statistics of Land Use Type in Jiading and Baoshan District

| Variables | Description | Percentage |
|---|---|---|
| Land use type | 0-industrial land | 18.3% |
| | 1-commercial land | 22.3% |
| | 2-educational land | 7.43% |
| | 3-technical land | 8.42% |
| | 4-residential land | 20.8% |
| | 5-greenspace land | 8.42% |
| | 6-agricultural land | 14.4% |

Table 2. Descriptive Statistics of Crash and interpretation variables at TAZ-level

| Variables | Mean | Min | Max | SD |
|---|---|---|---|---|
| Crash | 109 | 9 | 357 | 69.8 |
| TAZ area (km$^2$) | 3.26 | 0.75 | 13.42 | 2.4 |
| Arterial length (km) | 3.13 | 0.34 | 7.51 | 2.05 |
| Access density | 2.08 | 0.47 | 7.73 | 1.31 |
| Signal density | 1.74 | 0.543 | 4.052 | 0.746 |
| Road density (km/ km$^2$) | 3.107 | 0.287 | 12.627 | 2.13 |

## 3.2 Crash data

In this study, suburban arterials included national road, provincial road, first-class highway and second-class highway. All the suburban arterials of Jiading and Baoshan District were shown as Figure 1. There were 4 TAZs excluded since they didn't contain any arterials. All the modeling data were collected from 198 TAZs. Crash data in 2012 occurred on the arterials within each TAZ were collected. Firstly, the geocoding procedure in ArcGIS® was used to locate the crashes on the GIS base map, assistant with manually checking while without coordinate information.  Then overlay the TAZ shape on the base map and then calculate the crash numbers occurred on the arterials of each TAZ.

## 4 METHODOLOGY

### 4.1 Bayesian CAR model

In this study, a new strategy is proposed to analyze the safety impacts of effect factors on crash occurrence of arterials at TAZ level. The crash frequencies of arterials within each TAZ are used as dependent variables, aiming to analyze the full impacts of signal density and access density on the arterials.

The Bayesian modeling approach views parameters as random variables that are characterized by a prior distribution, and are estimated after combining the prior distribution and the sample data. The theoretical framework for Bayesian inference can be expressed as outlined in Equation 1:

$$\pi(\boldsymbol{\theta}|\boldsymbol{y}) = \frac{L(y|\theta)\pi(\theta)}{\int L(y|\theta)\pi(\theta)d\boldsymbol{\theta}} \tag{1}$$

where $y$ is the vector of observed data, $\theta$ is the vector of parameters required for the likelihood function, $L(y|\theta)$ is the likelihood function, $\pi(\theta)$ is the prior distribution of $\theta$, $\int L(y|\theta)\pi(\theta)d\theta$ is the marginal distribution of observed data, and $\pi(\theta|y)$ is the posterior distribution of $\theta$ given $y$.

In view of the existence of similar characters of traffic and road network among neighboring zones which could result in the spatial correlation between the crash occurrences on arterials in neighboring zones. Negative Binomial CAR model has been proposed in the Bayesian framework for this study to modeling for the crashes occurred on the arterials at TAZ-level. Let $y_i$ represent the number of crashes occurring within the $TAZ_i$, it assumes the dependent variable $y_i$ follows the Negative Binomial distribution as outlined in Equation 2:

$$y_i \sim Negbin(\lambda_i, k) \tag{2}$$

where $\lambda_i$ is the expectation of $y_i$, $k$ is the over-dispersion coefficient.

In this study, a random effect term $\phi_i$ was introduced into the above negative binomial model to develop the CAR model in Bayesian framework, in order to explain the spatial correlation between the neighboring TAZs. The model was outlined in Equation 3:

$$\log(\lambda_i) = \psi_i = X'\beta + \phi_i \tag{3}$$

where $X'$ is the covariate matrix, $\beta$ is the vector of regression coefficients.

The conditional distribution of the CAR prior distribution is defined as outlined in Equation 4:

$$\phi_i|\phi_{(-i)} \sim N(\sum_j \frac{w_{i,j}}{w_{i+}}\phi_j, \frac{1}{\tau_c w_{i+}}) \tag{4}$$

where $\phi_{-i}$ is collection of all $\phi$ except for $\phi_i$, $\tau_c$ is a precision parameter, $w_{i+}$ is the sum of $w_{i,j}$ in the TAZs that is adjacent to $TAZ_i$ and the joint prior distribution for $\phi_i$ is calculated according to Equation 5:

$$\pi(\varphi) \propto exp\left\{-\frac{\tau_c}{2}\sum_{i \neq j} w_{i,j}(\phi_i - \phi_j)^2\right\} \tag{5}$$

where π(φ) is the joint distribution of φ, and α represents that the likelihood function is proportional to the right-hand side of the equation.

The weight matrix $W$ with the entry $W_{i,j}$ of total lane number of those arterials connecting the neighboring TAZs is defined as outlined in Equation 7:

$$W_{i,j} = \begin{cases} n_{ij}, & \text{if } TAZ_i \text{ and } TAZ_j \text{ are adjacent} \\ 0, & \text{if } TAZ_i \text{ and } TAZ_j \text{ are not adjacent} \end{cases} \tag{7}$$

## 5 MODELING RESULTS

In this study Bayesian CAR models were developed and using the WinBUGS [8] to calibrate the model coefficients. Without reliable prior information, all the coefficients were assumed following the normal distribution $N(0, 10^{-5})$, the over-dispersion coefficient $k \sim Inverse - Gamma(10^{-3}, 10^{-3})$. In the process of developing models, 20000 iterations of one MCMC chain were run, and the first 2000 samples were discarded as the burn-in samples. The results of the model estimation were extracted after convergence was made, they are summarized in

Table 3.

Table 3. Posterior Summary of Parameter Estimation Results

| Variables | Mean (SD) | 95% BCI |
|---|---|---|
| Intercept | **2.352 (0.117)** | **(2.117, 2.579)** |
| Arterial length | **0.193 (0.014)** | **(0.166, 0.219)** |
| Access density | **0.108 (0.02)** | **(0.068, 0.146)** |
| Signal density | **0.359 (0.038)** | **(0.285, 0.432)** |
| Betweenness centrality | **1.705 (0.256)** | **(1.204, 2.21)** |
| Road density | **-0.031 (0.012)** | **(-0.054, -0.008)** |
| **Land use type** | | |
|   **Base**: Industrial land | | |
|   Commercial land | 0.136 (0.073) | (-0.01, 0.282) |
|   Educational land | -0.001 (0.101) | (-0.199, 0.201) |
|   Technical land | -0.119 (0.095) | (-0.304, 0.073) |
|   Residential land | **0.184 (0.072)** | **(0.043, 0.326)** |
|   Greenspace land | 0.022 (0.094) | (-0.161, 0.208) |
|   Agricultural land | 0.075 (0.08) | (-0.082, 0.231) |
| CAR effects | **0.205 (0.164)** | **(0.029, 0.656)** |

The coefficients in bold represent the significance of its estimation. The significant CAR effects confirms the existence of spatial correlation in crash occurrence among the neighboring TAZs. The significant variables were analyzed in the following section.

## 6 VARIABLE INTERPRETATION

The significant variables mainly including arterial length, access density, signal density and land use type.

Arterial length was found to be positively correlated with crash occurrence on the suburban arterials within the TAZ. It can be explained as that the exposure of vehicles running on the arterials may increase with the longer arterials, this tend to increase the crash occurrence. This finding is consistent with previous studies [1].

The coefficients of access density is significantly positive. This implies that the higher access density will increase the frequency of crash occurrence. It can be explained as that the frequently vehicles access to the main road through the access along the arterials, will disturb the traffic and tend to increase the likelihood of crash occurrence. There are several studies analyzed the safety effect of access density and similar conclusions have been reached. Wang et al. [5] found the higher access density tend to increase the likelihood of crash occurrence on suburban arterials.

Signal density was found to have positive effects on crash occurrence. It means that the higher density of signalized intersection leads to more crashes. This can be explained as that a confused traffic flow may appears in the condition of higher signal density, and more behavior of lane-changing and overtaking tend to increase the risk of collision. It is also consistent with the previous studies [5].

The coefficient for road density was significantly negative. This implies the higher road density tends to reduce the frequency of crash occurrence. This can be explained as that the higher road density may reduce the disturbance of local traffic to arterials since its thorough network of collector roads and branches, and the traffic management and facilities must be better than the one with lower road density. As a result, the higher road density may lead to less crashes occurring on arterials. Wang et al. [7] found that the state-maintained roads density have a positive effects on on-system crash occurrence.

The variable of land use type (residential land vs industrial stored land) was found to be significant in the model. The result showed that more crashes occurred on residential land compared with industrial stored land. It can be explained as that the traffic volume of arterials in residential land must be larger than industrial land, and this will lead to the higher probability of collision between vehicles running on the arterials, increasing the likelihood of crash occurrence. At the same time, the numbers of access on arterials in residential land may be larger than industrial stored land, this will increase the lateral disturbance to arterials. This is similar to the conclusion of previous research [6].

## 7 CONCLUSION AND DISCUSSION

This study attempts to identify the full impacts of access and signal density on crash occurrence on arterials at TAZ-level. In this way, the full impacts on arterial safety, including road segments and intersections, could be analyzed. Bayesian negative binomial CAR models were developed in this study, crashes occurred on the arterials within each TAZ were selected as response variable.

According to the estimation results, the significant CAR effects confirm the existence of cross-zonal spatial correlation in crash occurrence. Besides, five variables (i.e., arterial length, access density, signal density, road density and residential land vs industrial land) have statistically significant impacts on crash occurrence. Specifically, higher access density and signal density tend to increase the likelihood of crash occurrence, and dense road network is found to be correlated with less crashes. Compared with the other types of land use, residential land is positively associated with crash occurrence.

In the light of the above results, these models developed for the suburban arterials in this study appear to be useful with many applications such as a guideline of access and signal density for suburban arterials. At last, the spacing setting of access and signalized intersection need further investigation.